\documentclass[usenatbib]{mn2e}
\usepackage{graphics,epsfig,color}
\usepackage{amsmath}
\usepackage{amssymb}
\usepackage{url}
\usepackage{txfonts}

\usepackage{color}
\usepackage{natbib}
\usepackage{tabularx}

\def\degr{\hbox{$^\circ$}}

\def\arcsec{\hbox{$^{\prime\prime}$}}

\def\farcm{\hbox{$.\mkern-4mu^\prime$}}

\newcommand{\nh}{$N_\mathrm{H}$}

\newcommand{\ltsima}{$\buildrel < \over \sim$}
\newcommand{\lsim}{\lower.5ex\hbox{\ltsima}}
\newcommand{\gtsima}{$\buildrel > \over \sim$}
\newcommand{\gsim}{\lower.5ex\hbox{\gtsima}}

\newcommand{\hj}{HESS\,J1507$-$622}
\newcommand{\chandra}{\emph{Chandra}}
\newcommand{\xmm}{\emph{XMM-Newton}}
\newcommand{\suzaku}{\emph{Suzaku}}

\newcommand{\csrcTwo}{CXOU\,J150706.0$-$621443}
\newcommand{\csrcOne}{CXOU\,J150850.6$-$621018}
\newcommand{\ssrcTwo}{Suzaku\,J1507.0$-$6214}
\newcommand{\ssrcOne}{Suzaku\,J1508.8$-$6210}

\title[\suzaku\ observations of \hj]{Exploring the potential X-ray counterpart of the puzzling TeV gamma-ray source \hj\ with new \suzaku\ observations}

\author[P.~Eger, W.~F.~Domainko, J.~Hahn]
  {P.~Eger$^1$, W.~F.~Domainko$^1$, J.~Hahn$^1$, \\
    Max-Planck-Institut f\"ur Kernphysik, P.O. Box 103980, D-69029 Heidelberg, Germany}

\date{}

\pagerange{\pageref{firstpage}--\pageref{lastpage}} \pubyear{20xx}

\def\LaTeX{L\kern-.36em\raise.3ex\hbox{a}\kern-.15em
    T\kern-.1667em\lower.7ex\hbox{E}\kern-.125emX}

\begin{document}

\label{firstpage}

\maketitle

\begin{abstract}
The unidentified VHE (E$>$100\,GeV) gamma-ray source \hj\ seems to not fit into standard models for sources related to young supernova remnants, pulsar wind nebulae, or young stellar populations in general. 
This is due to its intrinsically extended, but yet compact morphology, coupled with a relative large offset ($\sim$3.5$^\circ$) from the Galactic plane.
Therefore, it has been suggested that this object may be the first representative of a new distinct class of extended off-plane gamma-ray sources.
The distance to \hj\ is the key parameter to constrain the source's most
important properties, such as age and energetics of the relativistic particle population. 

In this article we report on results of follow-up observations of the potential X-ray counterpart with \suzaku . 
We present detailed measurements of its spectral parameters and find a high absorbing hydrogen column density, compatible with the total amount of Galactic gas in this direction. 
In comparisons to measurements and models of the Galactic three-dimensional gas distribution we show that the potential X-ray counterpart of \hj\ may be located at the far end of the Galaxy.
If the gamma-ray source is indeed physically connected to this extended X-ray source, this in turn would place the object outside of the usual distribution of Galactic VHE gamma-ray emitters.
\end{abstract}


\section{Introduction}
\label{sec-introduction}
Multi-wavelength (MWL) observations are powerful tools to investigate the
nature of unidentified very-high-energy (VHE, E$>$100\,GeV) gamma-ray sources detected by ground-based imaging atmospheric Cherenkov telescopes such as H.E.S.S., VERITAS or MAGIC \citep[for a review, see e.g.][]{hinton2009}.
In particular, non-thermal X-ray sources are excellent observational tracers for
highly energetic particles, and are in many cases linked to VHE gamma-ray sources. 
Prominent examples of such cases are the supernova remnants (SNRs) Tycho \citep{decourchelle2001} and RX\,J1713.7$-$3946 \citep{cassam2004} as well as pulsar wind nebulae \citep[PWNe, for a review, see][]{gaensler2006}, such as MSH\,15-52 \citep{trussoni1996}. 
Two well-known Galactic VHE gamma-ray sources related to strong X-ray emitters are the PWNe HESS\,J1825$-$137 \citep{aharonian2005a,aharonian2006a} and Vela\,X \citep{aharonian2006b}.

One, typically unknown, key quantity of unidentified gamma-ray sources is their distance.
For Galactic sources, observations of their X-ray counterpart may help to
constrain their location in the Galaxy. 
Soft X-rays are absorbed by atomic and molecular interstellar gas along the line of sight, and from the level of this attenuation the total column density (CD) of traversed gas (predominantly hydrogen) can be estimated \citep[see e.g.][]{wilms2000}. 
If the distribution of atomic and molecular hydrogen in the Galaxy is known \citep[e.g.][]{dickey1990,ferriere2001,kalberla2005}, constraints on the distance of the source can be placed, based on X-ray spectral measurements. 

In the case where the level of soft X-ray attenuation is comparable to the total Galactic neutral hydrogen CD in this direction, the source is likely located at the far end or even outside of the Galaxy. 
This was the case for the VHE gamma-ray source HESS\,J1943$+$213, where the highly absorbed X-ray counterpart pointed towards an Extragalactic origin \citep{abramowski2011}. 
For sources with a significant angular offset from the Galactic plane, this method can also be used to constrain their physical distance from the disk.

In observations performed by the H.E.S.S. array of imaging atmospheric Cherenkov telescopes an extended VHE gamma-ray source with an angular off-set of $\sim$3.5$^{\circ}$ from the Galactic plane was discovered \citep[\hj ;][]{acero2011}. 
This source features the second largest angular offset (next to the nearby SNR SN\,1006) from the Galactic plane among all VHE gamma-ray sources that or not clearly linked to known Extragalactic objects. 
Despite detailed MWL analyses and theoretical considerations, the nature of this source has not been clearly identified yet. 
Interestingly, a faint, diffuse X-ray counterpart is likely connected to \hj\ (see sect.~\ref{sec-previous-observations}). 
The current literature favors a PWN interpretation for this object where the X-ray and VHE gamma-ray emission is produced by ultra-relativistic electrons through synchrotron and inverse-Compton radiation, respectively (see sect.~\ref{sec-discussion}). 
However, the distance to \hj , and its potentially associated X-ray nebula, is still unknown, but constitutes a key component for all attempts to model the evolution of the source as well as the radiation mechanisms. 

In this article we present detailed spectroscopic results for this X-ray counterpart, based on newly available \suzaku\ observations, which can provide a direct estimate of the total CD of hydrogen (\nh ) in the line of sight towards this source and thus of its distance.

\section{Previous X-ray observations with XMM-Newton and Chandra}
\label{sec-previous-observations}
Following up on its original detection by H.E.S.S., the region around \hj\ has been observed once with \chandra\ \citep{acero2011} and twice with \xmm\ \citep{tibolla2014}. 

In the 20\,ks \chandra\ observation (ObsID: 9975) several point-like sources were detected but classified to be most likely unrelated to \hj .
This \chandra\ observation also revealed two extended sources. 
One of them (\csrcOne ) is rather bright, with a flux of $F_\mathrm{X}$(2-10\,keV) = $7.0\pm 0.7\times 10^{-13}$\,erg\,cm$^{-2}$\,s$^{-1}$, but outside of the intrinsic VHE gamma-ray size of \hj\ \citep{acero2011}. 
As briefly outlined by \citet{tibolla2014}, assuming the X-ray source \csrcOne\ and the VHE gamma-ray source \hj\ are physically related, the lack of overlap between the two would be hard to explain, even in an aged PWN scenario. 
For such an object, despite a much more extended IC/TeV nebula compared to the  synchrotron/X-ray nebula, one still would expect IC emission from freshly injected electrons to overlap with the X-ray nebula, which is not seen when comparing \csrcOne\ and \hj . 
Prime examples for such aged PWN scenarios are HESS\,J1825$-$137 \citep{aharonian2005a} and HESS\,J1303$-$631 \citep{aharonian2005b}, where the TeV sources are much more extended but still partly overlap with the more compact X-ray nebulae (see \citet{uchiyama2009} and \citet{abramowski2012}, respectively). 
A link between \hj\ and the offset X-ray source \csrcOne\ might thus require an even more evolved, relic PWN scenario.

The second extended X-ray source (\csrcTwo ) was detected with a statistical significance of $\sim$7\,$\sigma$ by \chandra\ and is spatially consistent with the VHE gamma-ray emission region. 
This source is rather faint, with a flux of $F_\mathrm{X}$(2-10\,keV) = $1.1^{+0.3}_{-0.5}\times 10^{-13}$\,erg\,cm$^{-2}$\,s$^{-1}$, estimated from the \chandra\ count rate \citep{acero2011}. 
Given the low flux, the limited statistics from the 20\,ks \chandra\ observation did not allow for a detailed spectral study of \csrcTwo . 
The extensions of the sources were estimated to be 20-25\arcsec\ for \csrcTwo\ and 35-40\arcsec\ for \csrcOne , respectively \citep{acero2011}. 
Due to its positional coincidence and extended nature \csrcTwo\ was suggested as a potential counterpart to \hj\ \citep[see][]{acero2011}. 
To account for the very low X-ray flux compared to the flux in TeV gamma-rays, these authors suggested a relic PWN scenario with a very low magnetic field of $\sim$0.5$\mu G$. 
Such a low magnetic field leads to a very faint X-ray nebula along with a large accumulation of highly energetic electrons radiating predominantly TeV gamma-rays via the IC mechanism.

Unfortunately, both observations performed with \xmm\ (ObsIDs: 0556310201, 0651620101) suffered from long periods of strong background flaring activity, rendering a significant fraction of the exposure time unusable for scientific analyses \citep[see][]{acero2011,tibolla2014}. 
Even the remaining observation time of $\lesssim$10\,ks featured an increased background level which affected the overall sensitivity, particularly for extended sources. 
However, both of the above mentioned extended \chandra\ sources were also detected in the second \xmm\ observation \citep{tibolla2014}. 
\csrcTwo , the fainter of the two, was barely above the detection threshold with a statistical significance of $\sim$4\,$\sigma$. 
Again, due to the limited number of counts no detailed spectral analysis was possible for \csrcTwo . 
An additional source \citep[XMMU\,J150835.7$-$621021;][]{tibolla2014}, in close proximity (112$^{\prime\prime}$) to \csrcOne , was detected in the second \xmm\ observation but not seen in the earlier \chandra\ data despite being above the detection threshold if assuming constant flux.  
Therefore, \citet{tibolla2014} concluded that this source is probably variable and propose an X-ray binary or a flaring star as the the most likely scenarios.

\section{\suzaku\ data analysis}
\label{sec-analysis}
In September, 2012, the region of \hj\ was observed with \suzaku\ with two deep pointings, one of them (sequence No. 507025010, 79.8\,ks) centered on \hj\, the other (sequence No. 507026010, 40.9\,ks) centered on \csrcOne\ (see Tab.\,\ref{tab-observations} for more details).
We analyzed the XIS data using the most recent versions of the HEADAS software package \citep[v6.15.1;][]{blackburn1995}, and the HEASARC calibration database. 
We extracted images and spectra with \texttt{xselect} from the \texttt{cleaned} event lists using the recommended selection criteria (\texttt{STATUS$<$524287 \&\& (STATUS\%(2**17)$<$2**16)}) to remove events from the $^{55}$Fe calibration source. 

\begin{table}
\caption[]{\suzaku\ observations}
\begin{center}
\begin{tabular}{lllll}
\hline\hline\noalign{\smallskip}
\multicolumn{1}{l}{No.} &
\multicolumn{1}{l}{sequence} &
\multicolumn{1}{l}{exposure} &
\multicolumn{2}{c}{Pointing position} \\
\multicolumn{1}{l}{} &
\multicolumn{1}{l}{} &
\multicolumn{1}{l}{(ks)} &
\multicolumn{1}{c}{R.A.} &
\multicolumn{1}{c}{Dec.} \\
\noalign{\smallskip}\hline\noalign{\smallskip}
1 & 507025010 & 79.8 & 15:06:56.6 & $-$62:20:47 \\
2 & 507026010 & 40.9 & 15:08:43.6 & $-$62:09:52 \\
\hline\noalign{\smallskip}
\end{tabular}
\label{tab-observations}
\end{center}
\end{table}

\begin{figure}
  \begin{center}
  \resizebox{1.0\hsize}{!}{\includegraphics[clip=]{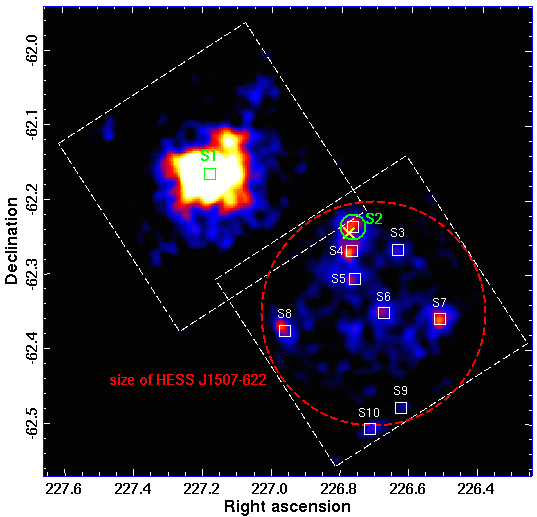}}  
  \caption{Combined mosaic of \suzaku\ XIS0 and XIS3 count maps from the two observations in the energy range 1-10\,keV, smoothed with a Gaussian kernel with a width of 0\farcm5. The dashed boxes (white) indicate the field of view of the two observations. The dashed circle (red) shows the position and intrinsic size of \hj . The small boxes show the positions of sources detected with \texttt{wavdetect}. The cross (yellow) indicates the position of the faint diffuse source detected with \chandra\ and \xmm , and the circle (green) shows the region used to extract the spectrum from \csrcTwo . The color scale is linear and adjusted such that the fainter sources are visible, but S1 is highly saturated.}
  \label{fig-mosaic}
  \end{center}
\end{figure}

We extracted count images from both front-illuminated detectors (XIS0 and XIS3) in the 1-10\,keV energy range. 
To generate images of the contribution from the non-X-ray background (NXB) in this energy range we used the tool \texttt{xisnxbgen} \citep{tawa2008}. 
After NXB subtraction we corrected the images for the mirror vignetting using a simulated exposure map created with the tool \texttt{xissim} \citep{ishi2009} at an energy of 2.5\,keV. 
Fig.~\ref{fig-mosaic} shows a combined mosaic image of the two \suzaku\ observations. 

To detect point-like and extended sources we used the \texttt{wavdetect} tool from the CIAO v4.6 software package \citep{fruscione2006}, setting the chance probability for a false detection to 1\%. 
The positions of all detected sources are listed in Tab.~\ref{tab-regions} and indicated by small squares in Fig.~\ref{fig-mosaic}. 
The table also lists likely counterparts from previous \chandra\ and \xmm\ observations \citep{acero2011,tibolla2014}. 
With the exception of two sources (Suzaku\,J1506.7$-$6221 and Suzaku\,J1506.5$-$6229) all detected objects have counterparts from previous X-ray observations (see Tab.~\ref{tab-regions}). 
However, due to their low count rates, no more detailed analyses concerning their spectra or extensions are possible. 

Due to the significantly larger PSF of \suzaku\ two previously detected X-ray sources fall within the range of \ssrcOne : The bright extended source \csrcOne\ \citep{acero2011}, also detected by \xmm\ with consistent flux  \citep[XMMU\,J150851.1$-$621017;][]{tibolla2014}, 
 and the much fainter but variable source XMMU\,J150835.7$-$621021 \citep[][see also sect.~\ref{sec-previous-observations}]{tibolla2014}. 
With \xmm\ the flux of XMMU\,J150835.7$-$621021 was detected at $\sim$10\% of the flux from \csrcOne . 
However, due to the apparent variability, the relative flux contributions may be different in this \suzaku\ observation (for further discussion see below). 
Due to this potential issue of source confusion, we will use the new \suzaku\ name for this source whenever we refer to the current analysis. 

\begin{figure}
  \begin{center}
  \resizebox{1.0\hsize}{!}{\includegraphics[clip=]{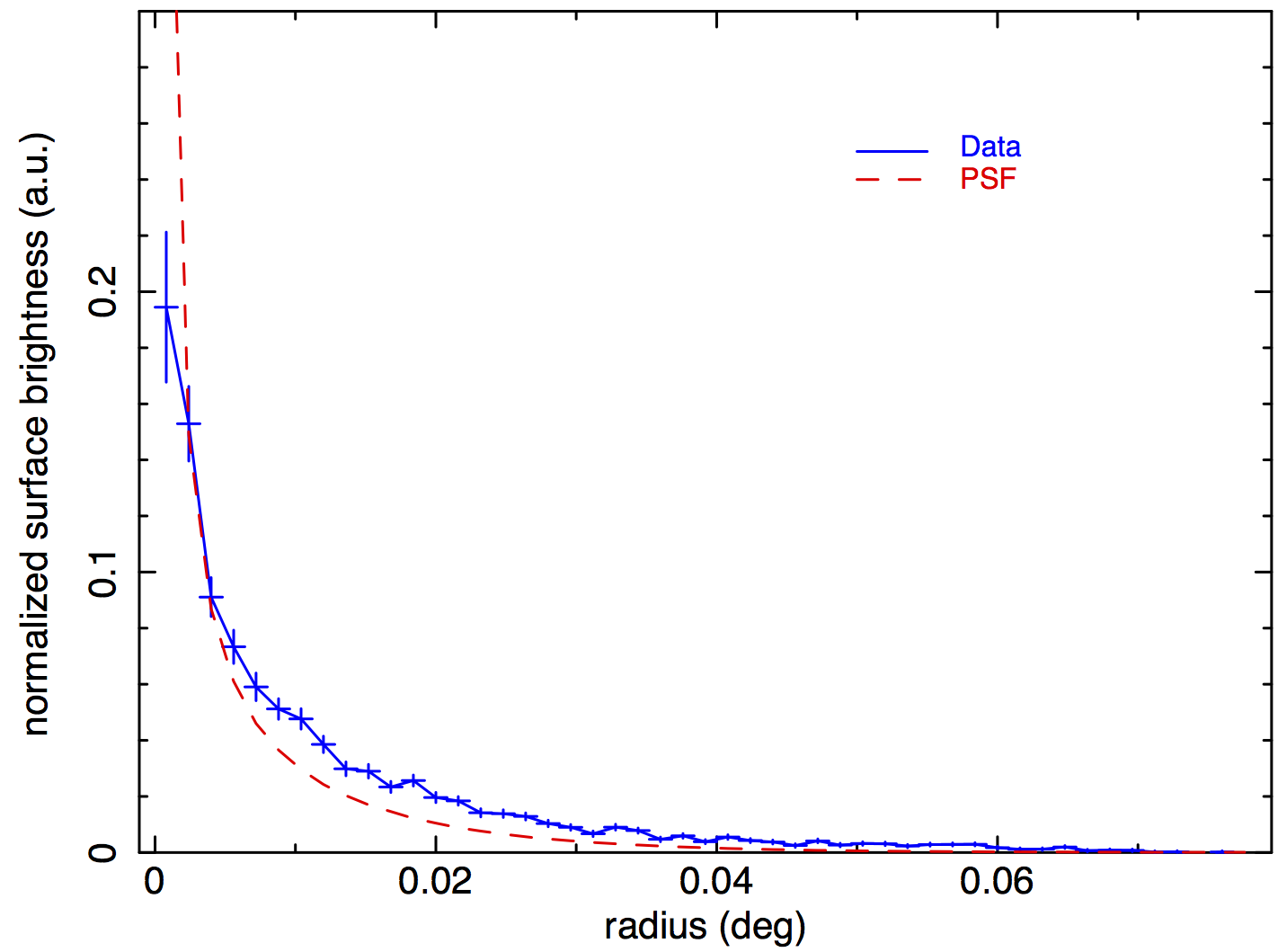}}  
  \caption{Radial profile of \ssrcOne\ (1-10\,keV) from the XIS0$+$3 counts map (markers with error bars, blue), and the \suzaku\ PSF at 4.5\,keV (dashed, red).}
  \label{fig-profile}
  \end{center}
\end{figure}

To test for an extension beyond the \suzaku\ PSF of \ssrcOne\ we extracted its radial profile from the unsmoothed XIS0$+$3 counts image (1-10\,keV). 
This profile is shown in Fig.~\ref{fig-profile} and compared to the on-axis PSF at an energy of 4.5\,keV as stored in the \suzaku\ calibration database. 
To account for the diffuse astrophysical background component we subtracted from the data the surface flux level measured at offsets larger than 0.07$^\circ$ from the source position. 
The relative normalization between the data and PSF profiles was calculated such that they yield the same integral between 0 and 0.06$^\circ$. 
As is evident from Fig.~\ref{fig-profile}, the morphology of \ssrcOne\ is incompatible with a point-like source, as already stated by \citet{matsumoto2014} and \citet{sakai2013}. 
However, due to the comparatively large PSF of \suzaku , it is unclear how much of this apparent extent is due to the intrinsic size of \csrcOne\ or due to confusion with the offset source XMMU\,J150835.7$-$621021 (see previous paragraph). 
Therefore, we suggest to refer to the extension measurements of \csrcOne\ with \chandra\ \citep{acero2011} and \xmm\ \citep{tibolla2014} for a more reliable estimate of this parameter. 

Of particular interest here is the faint extended source \csrcTwo , previously detected with \chandra\ and \xmm , which can be also clearly identified in the new \suzaku\ observation: \ssrcTwo .
The position of \ssrcTwo\ is slightly shifted towards the North-West compared to the \chandra\ and \xmm\ positions of \csrcTwo\ \citep[see][]{acero2011,tibolla2014}, as indicated by the yellow cross in Fig.~\ref{fig-mosaic}. 
However, the high-resolution \chandra\ results do not show any other source towards the shifted direction and we thus rule out source confusion due to \emph{Suzaku's} larger point spread function as the origin of the shift. 
More likely, the relative offset originates from uncertainties in the absolute pointing position of $\sim$20\arcsec\ \citep[see e.g.][]{uchiyama2008}. 
We see similar offsets between the \suzaku\ and \chandra\ positions also for the other detected sources with \chandra\ counterparts in this observation. 
We therefore identify \ssrcTwo\ with \csrcTwo\ and use the original \chandra\ name for this object for all further discussion in this paper.

We extracted spectra from all sources which offered a sufficient number of counts (i.e. S1, S2, S4, S7, and S8) from all three detectors (XIS0, XIS1, XIS3). 
For all sources but \ssrcOne\ we chose an extraction radius of 60\arcsec\ which is recommended for point-like source analyses, and a nearby source-free region for the extraction of the background spectrum.
Due to its intrinsic size and brightness we chose a larger radius of 120\arcsec\ for \ssrcOne\ and a large concentric annular region for the background (also see Tab.~\ref{tab-regions}). 
These new deep \suzaku\ data offer the opportunity to study the spectrum of the faint extended source \csrcTwo\ which was not possible with previous observations by \chandra\ and \xmm\ due to the limited count statistics.

\begin{table*}
\caption[]{Detected sources}
\begin{center}
\begin{tabular}{lllllll}
\hline\hline\noalign{\smallskip}
\multicolumn{1}{l}{Source \#} &
\multicolumn{2}{c}{position} &
\multicolumn{1}{l}{Source name} &
\multicolumn{1}{l}{radius$^{(1)}$} &
\multicolumn{1}{l}{background region$^{(1)}$} &
\multicolumn{1}{l}{X-ray counterpart$^{(2)}$} \\
\multicolumn{1}{l}{} &
\multicolumn{1}{c}{R.A.} &
\multicolumn{1}{c}{Dec.} &
\multicolumn{1}{c}{Suzaku} &
\multicolumn{1}{l}{(\arcsec)} &
\multicolumn{1}{l}{} &
\multicolumn{1}{l}{} \\
\noalign{\smallskip}\hline\noalign{\smallskip}
S1 & 15:08:50.6 & $-$62:10:18 &J1508.8$-$6210 & 120 & annulus:  260\arcsec - 470\arcsec				& CXOU\,J150850.6$-$621018 \\  
   &                          &               & -	&    & & XMMU\,J150851.1-621017 \\
   &                          &               & -	&    & & XMMU\,J150835.7$-$621021 \\
S2 & 15:07:04.9 & $-$62:14:16 &J1507.0$-$6214 & 60  & offset, source-free region   & CXOU\,J150706.0$-$621443 \\  
S3 & 15:06:35.9 & $-$62:16:24 &J1506.6$-$6216 & -	& - 							& CXOU\,J150636.9$-$621628 \\
S4 & 15:07:08.2 & $-$62:16:33 &J1507.1$-$6216 & 60  & offset, source-free region   & CXOU\,J150708.8$-$621643 \\
   &                          &               & -	&    &  & XMMU\,J150708.4$-$621642 \\
S5 & 15:07:06.0 & $-$62:18:45 &J1507.1$-$6218 & -	& - 							& CXOU\,J150706.7$-$621858 \\
S6 & 15:06:45.2 & $-$62:21:30 &J1506.7$-$6221 & -	& - 							& - 					   \\
S7 & 15:06:06.3 & $-$62:21:58 &J1506.1$-$6221 & 60  & offset, source-free region   & CXOU\,J150606.7$-$622210 \\
S8 & 15:07:54.0 & $-$62:22:58 &J1507.9$-$6222 & 60  & offset, source-free region   & CXOU\,J150756.0$-$622238 \\
   &                          &               & -	&    &  & XMMU\,J150755.9$-$622237 \\
S9 & 15:06:32.9 & $-$62:29:07 &J1506.5$-$6229 & -	& - 							& - 					   \\
S10& 15:06:54.5 & $-$62:30:50 &J1506.9$-$6230 & -	& - 							& CXOU\,J150656.1$-$623040 \\
\hline\noalign{\smallskip}
\end{tabular}
\label{tab-regions}
\end{center}
$^{(1)}$ Spectral extraction radius and description of background region in the case the statistics allowed for spectral fitting. 
$^{(2)}$ Likely counterpart from previous observations with \xmm\ and \chandra\ \citep[see][]{acero2011,tibolla2014}. 
\end{table*}

We performed the spectral fits with XSPEC v12 \citep{arnaud1996} and used an absorbed powerlaw as well as a plasma (MEKAL) model to investigate both non-thermal and thermal radiation mechanisms. 
To model the photo-electric absorption we used the \texttt{tbabs} model along with the Galactic metal abundances from \citet{wilms2000}. 
For each source we fitted the spectra from all detectors simultaneously with linked spectral parameters. 
The spectral results are compiled in Tab.~\ref{tab-spectrum-results}. 

For four of these sources, flux estimates from previous X-ray observations with \xmm\ and \chandra\ are available, three of them based on spectral fitting. 
For Suzaku\,J1507.9$-$6222 a previous flux estimate is available only from \xmm\  where the flux was calculated from the count-rate assuming a certain spectral shape \citep[see][for details]{tibolla2014}. 
Here, the new \suzaku\ flux is a factor of $\sim$2 lower compared to \xmm , but still barely within the 1\,$\sigma$ statistical uncertainties. 
This discrepancy is probably due to differences between the assumed spectral parameters in the \xmm\ measurement and the measured spectral shape in this work. 
Suzaku\,J1507.1$-$6216 has an \xmm\ and \chandra\ counterpart with available spectral fitting results for \xmm\ \citep[see][]{tibolla2014}. 
The fluxes are compatible within 1\,$\sigma$ uncertainties, however, the best-fit spectral slopes deviate significantly. 
This is most likely due to the fact that in the \xmm\ analysis the absorption column density was fixed at the total Galactic value in this direction, whereas here we find that the source spectrum shows very little absorption with an upper limit on \nh\ well below the total Galactic value (see Tab.~\ref{tab-spectrum-results}). 
Correlations between \nh\ and the spectral slope likely explain the discrepancies in spectral index. 

\begin{figure*}
  \begin{center}
  \resizebox{1.0\hsize}{!}{\includegraphics[clip=]{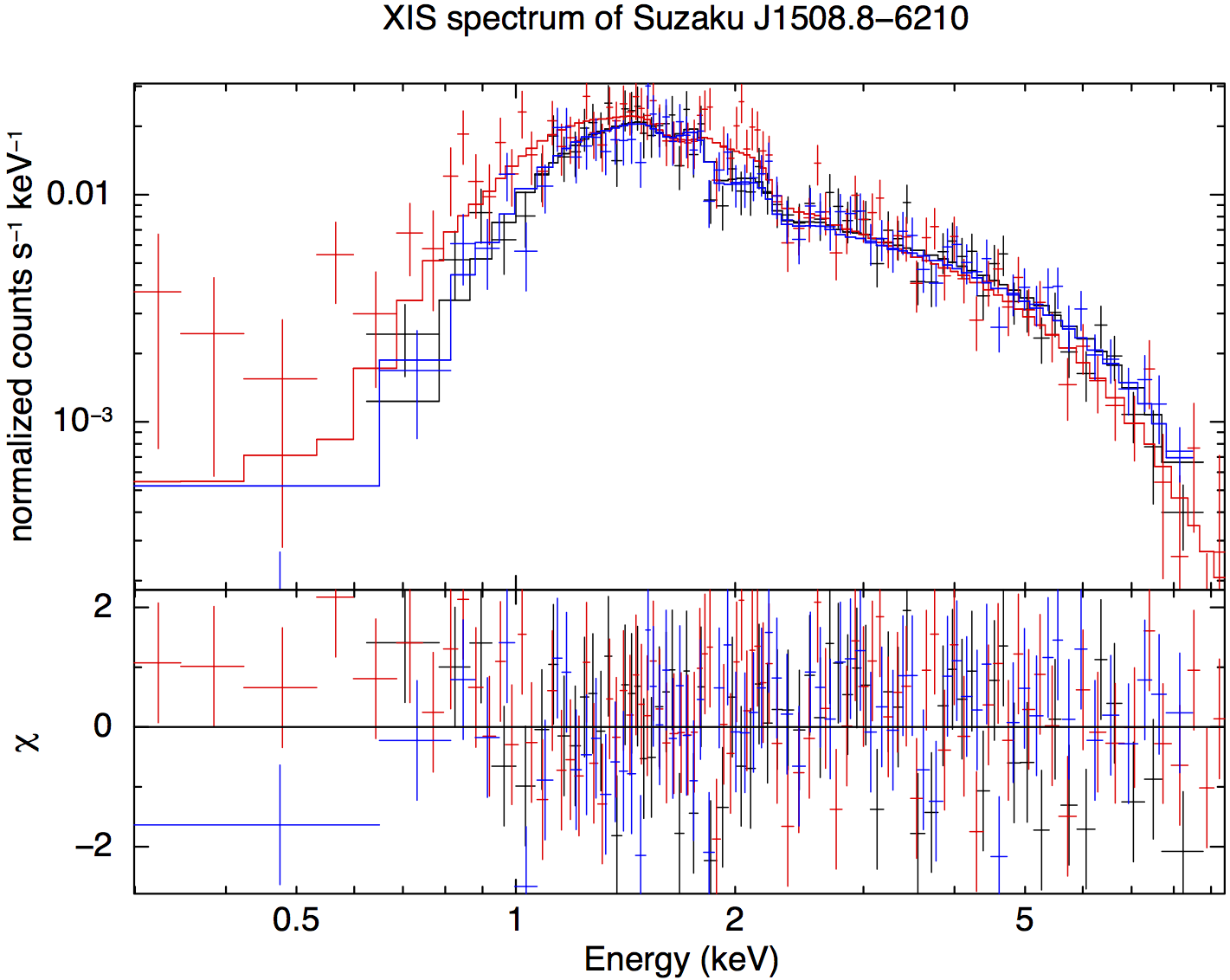}
  				\includegraphics[clip=]{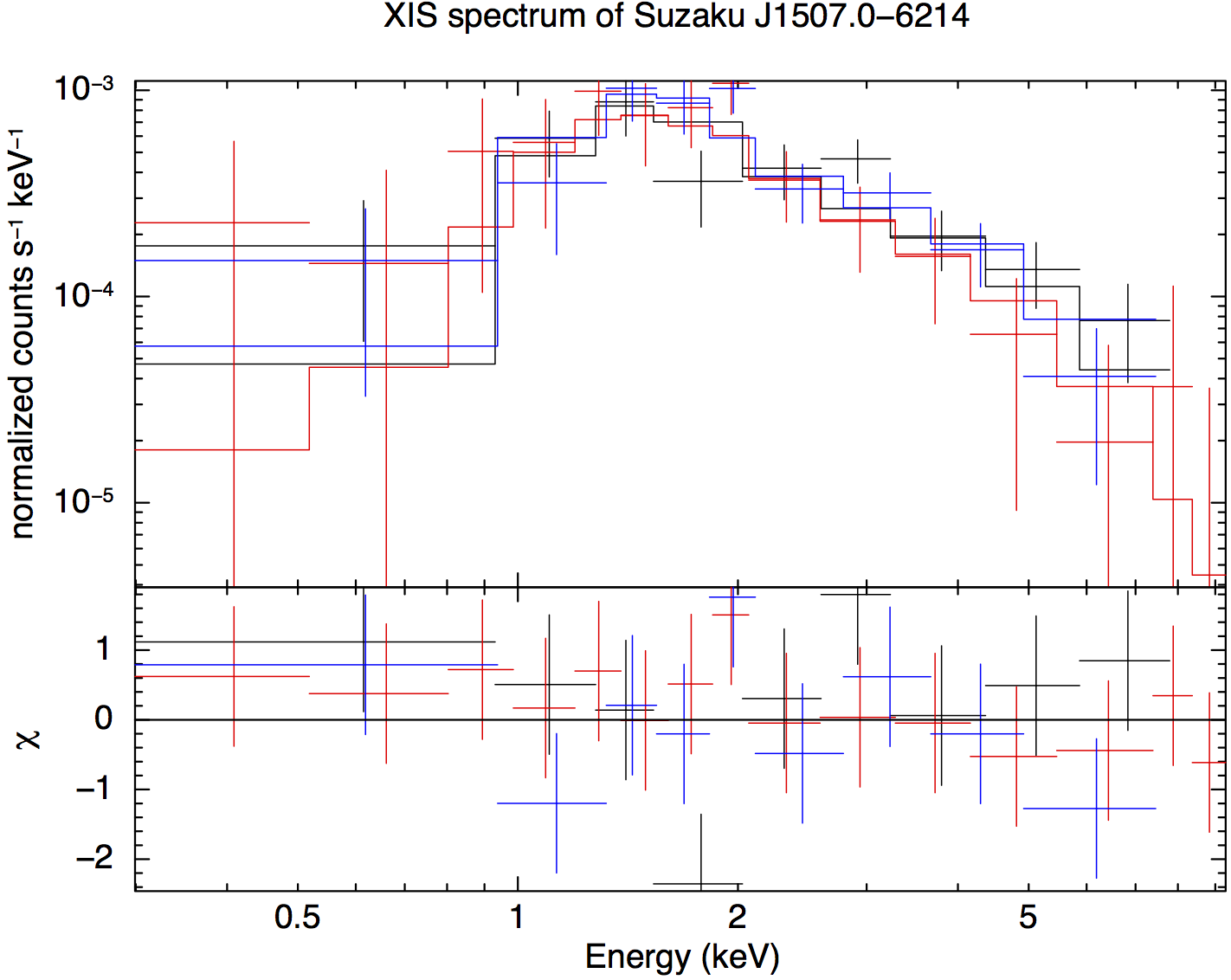}}
  \caption{\suzaku\ spectra of the two extended sources \ssrcOne\ (left) and \csrcTwo\ (right) shown together with the best-fit absorbed powerlaw model (stepped lines). The colors indicate the spectra from the individual cameras: XIS-0 (black), XIS-1 (red), XIS-3 (blue).}
  \label{fig-spectra}
  \end{center}
\end{figure*}

Below, we discuss the results for the two extended sources \ssrcOne\ and \csrcTwo\ in more detail. 
Figure~\ref{fig-spectra} shows the spectra for these two sources along with the best-fit powerlaw models. 
Judging from the $\chi^2$ values, the powerlaw model is clearly preferred for \ssrcOne\ ($\geq 5\,\sigma$), whereas there is no clear preference for \csrcTwo , given the current data. 
For both sources, the value of \nh\ appears to be large (see discussion in sect.~\ref{sec-nh-measurements}) and, in the case of \ssrcTwo , rather independent of the assumed spectral shape. 

\citet{matsumoto2014} and \citet{sakai2013} already reported about results for these two extended sources based on the same \suzaku\ dataset. 
These authors confirmed the extended nature of \ssrcOne . 
However, the lower count statistics from \csrcTwo , coupled with its more compact size, make it point-like for \suzaku . 
Our spectral results are compatible in terms of photon index and flux with the ones obtained by the authors above. 
However, we find systematically larger values for the hydrogen column density \nh . 
Unfortunately, the description of the spectral analysis by these authors is not complete, and key information like the used absorption model and the assumed metal abundances are missing. 
We were able to reproduce the results of \citet{matsumoto2014} and \citet{sakai2013} by changing from Galactic to solar metal abundances. 
Because of the higher metallicity of the sun compared to the average Galactic level, a lower equivalent hydrogen CD is needed for the same absorption effect, thus the lower \nh\ value in their fit. 
However, because these sources are very likely of extra-solar origin, we remain with Galactic abundances and deem our results to be more realistic. 

As already discussed in the first paragraph of this section, \ssrcOne\ may be composed of the two independent X-ray sources \csrcOne\ and XMMU\,J150835.7$-$621021. 
To compare with the previous \chandra\ result \citep{acero2011} we calculated a flux of \ssrcOne\ in the 2-10\,keV band of $(9.8\pm 0.3)\times 10^{-13}$\,erg\,cm$^{-2}$s$^{-1}$ which is about 30\% higher than the \chandra\ measurement of \csrcOne . 
This difference might arise from an additional contribution from the fainter but variable source XMMU\,J150835.7$-$621021. 
In this case the latter source must have had a $\sim$3 times higher flux during the \suzaku\ observation than during the earlier \xmm\ measurement \citep[see][]{tibolla2014}. 
Variability of this magnitude is very well plausible for flaring stars \citep[see, e.g.][]{liefke2010} and X-ray binaries \citep[for a review, see][]{vandenberg2010}, the two most plausible scenarios for XMMU\,J150835.7$-$621021 as suggested by \citet{tibolla2014}. 
Another contributing factor to the larger flux seen with \suzaku\ from \ssrcOne\ compared to \chandra\ and \xmm\ could arise from the larger region used for spectral extraction. 
However, this would only be the case if \csrcOne\ is much more extended than estimated from the \xmm\ and \chandra\ maps, with a surface flux in the tails below their detection limit.

\begin{table}
\caption[]{X-ray spectral fitting results}
\renewcommand{\tabcolsep}{4pt}
\begin{center}
\begin{tabular}{llllll}
\hline\hline\noalign{\smallskip}
\multicolumn{1}{l}{source} &
\multicolumn{1}{l}{\nh$^{(1)}$} &
\multicolumn{1}{l}{$\Gamma$ / $kT^{(2)}$} &
\multicolumn{1}{l}{$F_\mathrm{obs}^{(3)}$} &
\multicolumn{1}{l}{$F_\mathrm{unabs}^{(4)}$} &
\multicolumn{1}{l}{$\chi^2$ / ndf}\\[0.05cm]
\multicolumn{1}{l}{} &
\multicolumn{1}{l}{(10$^{21}$\,cm$^{-2}$)} &
\multicolumn{1}{l}{(-) / (keV)} &
\multicolumn{2}{c}{(10$^{-13}$\,erg\,cm$^{-2}$\,s$^{-1}$)} &
\multicolumn{1}{l}{} \\
\noalign{\smallskip}\hline\noalign{\smallskip}
\multicolumn{3}{l}{\emph{powerlaw}} \\
S1 & 7.4$^{+1.0}_{-1.0}$ & 1.83$^{+0.08}_{-0.08}$ & 11.9$^{+0.3}_{-0.2}$ & 13.8$^{+0.5}_{-0.5}$ & 222.19 / 223 \\[0.1cm]
S2 & 11.7$^{+8.8}_{-6.5}$ & 2.2$^{+0.6}_{-0.5}$ & 0.89$^{+0.14}_{-0.12}$ & 1.6$^{+0.3}_{-0.3}$ & 23.93 / 30 \\[0.1cm]
S4 & $<$1.7 & 2.5$^{+0.6}_{-0.4}$ & 0.55$^{+0.09}_{-0.08}$ & 0.58$^{+0.12}_{-0.10}$ & 52.1 / 50\\[0.1cm]
S7 & 26$^{+27}_{-15}$ & 2.0$^{+0.9}_{-0.7}$ & 1.1$^{+0.1}_{-0.1}$ & 2.1$^{+1.0}_{-0.9}$ & 54.7 / 45\\[0.1cm]
S8 & $<$15 & 1.7$^{+0.9}_{-0.8}$ & 1.0$^{+0.4}_{-0.3}$ & 1.4$^{+0.9}_{-0.7}$ & 30.4 / 38\\[0.1cm]
\noalign{\smallskip}\hline\noalign{\smallskip}
\multicolumn{3}{l}{\emph{MEKAL}} \\
S1 & 4.7$^{+0.6}_{-0.6}$ & 8.5$^{+1.4}_{-1.5}$ & 11.8$^{+0.4}_{-0.3}$ & 12.9$^{+0.4}_{-0.4}$ & 266.81 / 223 \\[0.1cm] 
S2 & 11.4$^{+7.5}_{-6.7}$ & 3.3$^{+3.7}_{-1.2}$ & 1.0$^{+0.3}_{-0.2}$ & 1.5$^{+0.3}_{-0.3}$ & 23.48 / 30 \\[0.1cm]
S4 & $<$0.7 & 2.8$^{+1.2}_{-0.7}$ & 0.57$^{+0.6}_{-0.07}$ & 0.58$^{+0.13}_{-0.12}$ & 59.8 / 50\\[0.1cm]
S7 & 25$^{+23}_{-14}$ & 4.6$^{+8.0}_{-2.2}$ & 1.1$^{+0.3}_{-0.2}$ & 1.8$^{+0.7}_{-0.5}$ &  54.4 / 45\\[0.1cm]
S8 & 11$^{+11}_{-9}$ & 6.5$^{+25}_{-3.7}$ & 1.0$^{+0.4}_{-0.2}$ & 1.3$^{+0.7}_{-0.6}$ & 29.7 / 38\\[0.1cm]
\hline\noalign{\smallskip}
\end{tabular}
\label{tab-spectrum-results}
\end{center}
all quoted uncertainties are at 68\% confidence, and calculated with all other model fit parameters free to vary;
$^{(1)}$absorbing hydrogen column density (CD); 
$^{(2)}$photon index or plasma temperature, depending on the assumed model; 
$^{(3)}$observed energy flux (1-10\,keV); 
$^{(4)}$unabsorbed energy flux (1-10\,keV), with the effects of foreground absorption removed
\end{table}

\section{Discussion}
\label{sec-discussion}
Out of the two extended X-ray sources in the vicinity of \hj , originally detected by \chandra , in the following discussion we deem \csrcTwo\ to be the more likely counterpart to the TeV emission due to its spatial coincidence with the intrinsic size of \hj , which is not the case for \csrcOne\ (\ssrcOne ). 
Due to the lack of overlap with the VHE gamma-ray size of \hj , the latter source was considered a less likely counterpart in a PWN scenario \citep[see][and also Sect.~\ref{sec-previous-observations}]{acero2011,tibolla2014}.

After the discovery of \hj , a leptonic scenario was favored \citep{acero2011} while hadronic scenarios were considered unlikely \citep{domainko2011}.
\citet{acero2013} reported a point-like \emph{Fermi-LAT} counterpart with soft spectrum. 
However, the nature of the source still remains elusive. 
On the one hand, several PWN models were successfully applied to describe the emission seen from \hj : \citet{tibolla2011} proposed an ancient PWN model \citep[developed by][]{dejager2009}, \citet{tibolla2012} fitted a modified leaky-box model of \citep{zhang2008}, and \citet{vorster2013} managed to describe \hj\ with a time-dependent PWN model. 
On the other hand, no pulsar has been detected in the vicinity of \hj\ \citep{acero2011}.

The nature of the TeV gamma-ray source \hj\ is challenging to explain with established models for Galactic sources, such as young supernova remnants and PWNe, due to its large offset from the Galactic plane coupled with its apparent compactness.
In a PWN scenario the latter feature may indicate either that the source is nearby and still young, or that the source is more evolved and far away.
The problems with the first scenario are the absence of a young ($\lesssim 10^4$\,yr) and powerful pulsar and the relatively low X-ray flux of the candidate synchrotron nebula \csrcTwo . 
To explain the low X-ray flux, in the second scenario the PWN would be a more evolved system 
\citep[e.g.][]{mattana2009,balbo2010}. 
In this framework \citet{acero2011} found a distance of \hj\ of $>$6\,kpc by comparing its angular extent with the size of the nearby, evolved Geminga PWN. 
A distance of 6\,kpc has also been adopted in previous PWN models \citep{tibolla2011,tibolla2012,vorster2013}.
If the extension of \hj\ is driven by diffusion of energetic electrons for a time-scale of $2\times 10^4$~years,
the distance to the object needs to be very large ($\gtrsim$10\,kpc) to be compatible with the apparent compactness of the TeV source \citep{domainko2012}. 
This in turn implies a large physical distance of \hj\ from the Galactic plane, in particular much larger than the scale height of the distribution of VHE gamma-ray emitting pulsars. 
A remedy for this situation would be an association to hyper-velocity stars or \hj\ being a member of a new population of gamma-ray sources associated to older stellar populations. 
For a more detailed discussion of these scenarios see \citet{domainko2012} and \citet{domainko2014}. 

As becomes evident from the above discussion, the most important, but yet unknown parameter is the distance of \hj\ from Earth. 
Also, the details of the spectral shape and the flux of the non-thermal X-ray counterpart to \hj\ are very valuable to constrain intrinsic source properties, such as magnetic field and age. 

In the subsections below we explore the implications of the measured spectral parameters of \csrcTwo\ on its distance (using \nh ), and on the properties of the underlying relativistic particle population (using the spectral index and flux).

\subsection{Modeling the spectral energy distribution}
\label{sec-sed-model}
\begin{figure}
  \begin{center}
  \resizebox{1.\hsize}{!}{\includegraphics[clip=]{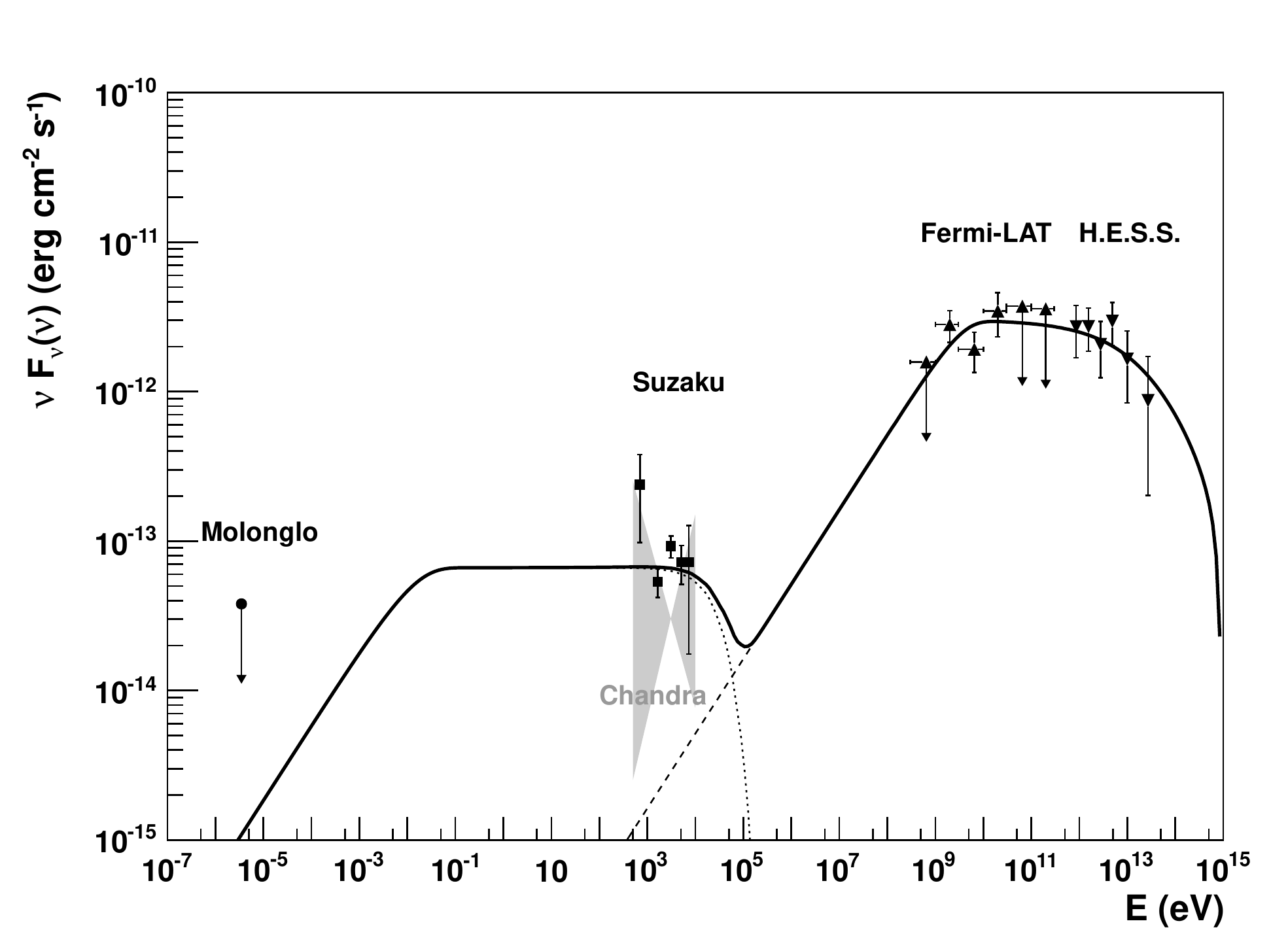}}  
  \caption{Broad-band SED of \hj , including the Radio flux upper limit from Molonglo \citep{domainko2012,bock1999}, the uncertainty band of the best-fit powerlaw spectrum measured with \chandra\ \citep[shaded area,][]{tibolla2014}, the \suzaku\ data points for \csrcTwo\  from this work, Fermi measurements in the high-energy gamma-ray regime \citep{domainko2012} and H.E.S.S. flux points at energies larger than 100\,GeV \citep{acero2011}. Additionally, the model SED is shown, assuming Synchrotron (dots) and IC emission (dashes) from a broken powerlaw electron distribution. See sect.~\ref{sec-sed-model} for a detailed description of the model.}
  \label{fig-SED}
  \end{center}
\end{figure}

We use our spectral measurement of \csrcTwo\ for an updated
modeling of the spectral energy distribution (SED) of \hj . 
We assume that \csrcTwo\ is physically related to \hj\ because of its extended nature and its spatial coincidence with the VHE gamma-ray source. 
Fig.~\ref{fig-SED} shows the broad-band spectral measurements used for this study. 
Here, the X-ray data points are the unfolded (corrected for instrument response) data from the \suzaku\ observation, corrected for the foreground absorption as determined by the best-fit powerlaw model, taking also parameter uncertainties into account (hence the large error bars of the lowest energy data point where the influence of \nh\ is strongest). 
In addition to the new \suzaku\ spectral data points of \csrcTwo\ we show in Fig.~\ref{fig-SED} also the uncertainty range of the previous best-fit powerlaw model measured with \chandra\ \citep[see][]{tibolla2014}. 
The latter is only an approximation, calculated from the quoted uncertainties of the integral flux and photon index, assuming no parameter correlations (as the full covariance matrix of the \chandra\ fit has not been published). 
The uncertainties of the \chandra\ measurement are very large due to limited statistics and both spectra are compatible within errors. 
The slightly lower flux seen with \chandra\ may arise from the smaller size of the spectral extraction region compared to the new \suzaku\ analysis. 

A case for a leptonic origin of \hj\ has been made in the past by various authors \citep{acero2011,tibolla2011,tibolla2012,vorster2013} in the framework of a PWN scenario \citep[see also][]{weiler1978,gaensler2006}.
One-component broken powerlaw electron spectra \citep{torres2014} and
two-component electron spectra \citep{vorster2013} were adopted to fit the SED of several PWNe. 
For \hj , using the flux measurement from the 2-year \emph{Fermi-LAT} source catalog (2FGL) below 100\,GeV \citep{nolan2012}, \citet{vorster2013} were able to fit the SED of \hj\ with a two-component electron spectrum. 
Here, we adopt the result of the longer 34-month \emph{Fermi-LAT} data set analyzed by \citep{domainko2012} for energies below 100\,GeV, following the approach of \citet{torres2014}, and fit a one-component electron spectrum. 
This is mainly done to reduce the number of free parameters compared to a two-component injection spectrum.

We assume a broken powerlaw energy distribution for the 
radiating electrons with a spectral break of $\Delta \Gamma = 1$, as expected
for a cooling break for continuous injection over longer periods of time caused by synchrotron and inverse-Compton radiation in the Thomson regime. 
With these assumptions we find a best-fit model with a spectral index below the break energy ($E < E_\mathrm{break}$) of $\Gamma = 2$, for $E > E_\mathrm{break}$ a spectral index of $\Gamma = 3$, and $E_\mathrm{break}$ = 0.9\,TeV.
The total energy in electrons is $5 \times 10^{47} (d/ 1\, \mathrm{kpc})^2$\,erg
with $d$ being the distance to the source. 
The maximum energy of electrons in this model is $E_\mathrm{max}$ = 1\,PeV, a value which is necessary to reproduce the highest energy X-ray data points.

From this model fit, constraints on the age of the radiating electrons can be obtained.
For the case were the break at 0.9\,TeV in the broken powerlaw is introduced by inverse-Compton cooling on the CMB, this would point towards an age of the source of $\approx 10^6$~years, i.e. a rather old source. 
This can be compared to age estimates based on the evolution of PWNe if such an origin is adopted for \hj .
\citet{mattana2009} found that the ratio of the flux in VHE gamma-rays (1-10\,TeV) $F_\gamma$ and the flux in X-rays
(2-10\,keV) $F_\mathrm{X}$ strongly increases with PWN age. 
Applying this method for \hj\ with log$\left(F_\gamma/F_\mathrm{X}\right) \approx 1.85$, an age of about $3 \times 10^4$\,years and thus a rather evolved PWN would be found. 
This is consistent with the fact that the size of the TeV source is much larger than the size of the X-ray source for evolved PWN \citep{kargaltsev2010} which also appears to be the case for \hj .
An age of about $3 \times 10^4$\,years would be in line with the PWN models of \citet{tibolla2012} and \citet{vorster2013}.
This age estimate is significantly smaller than the age of the Geminga PWN. Therefore, in the PWN scenario, \hj\ could be significantly
less extended than the Geminga PWN and could thus be located closer than the 6\,kpc as discussed above
(note that the multi-kpc distance estimate in Sect.~\ref{sec-distance} is independent from spectral modelling and only based on the X-ray absorption measurement of the potential X-ray counterpart).
The caveat for this interpretation is the smaller $F_\gamma/F_\mathrm{X}$ ratio of Geminga with respect to \hj . 
\citet{vorster2013} proposed the passage of the reverse shock to resolve this discrepancy.
In this paper we did not consider the effect of a past reverse shock passage on \hj . 
Our modelling parameters are given for the present particle spectrum in this source.
It has also to be noted that no supernova remnant has been detected around \hj .
Additional constraints on the age of radiating electrons can be placed by the maximum particle energy. 
Electrons with an energy of about 1\,PeV cool very fast via inverse-Compton radiation, and have to have been injected less than 1\,kyr ago into the system.
This may indicate the presence of a second spectral component at the highest energies and more generally may indicate that a two-component electron spectrum seems to be favored over a one-component electron spectrum.
To summarize, for a leptonic scenario the SED of \hj\ suggests a rather evolved system albeit with recent injection of highly energetic particles. 

In this model the magnetic field $B$ is 0.47 $\mu$G. 
This value for $B$ is consistent with the estimates of by \citep{acero2011} but is smaller than the values found by \citep[][1 $\mu$G]{domainko2012} and
\citep[][1.7 $\mu$G]{vorster2013}. 
The reason for this difference is the fact that here the measured hard X-ray spectrum for \csrcTwo\ is used for the SED modeling, in contrast to the softer spectra assumed in the previous studies. 

One caveat of such leptonic one-zone models is the significant difference of sizes of the emission regions in the X-ray and gamma-ray regimes. 
This may introduce a bias (i.e. underestimation of the synchrotron flux) in models where it is assumed that the X-ray and TeV fluxes are produced by the same population of electrons. 
With \chandra\ the size of \csrcTwo\ was estimated to be 20-25\arcsec\ \citep{acero2011}, and with the new \suzaku\ data the associated source (\ssrcTwo ) appears to be point-like and its flux to be contained within a radius of $\sim$60\arcsec\ (which is also the suggested spectral extraction radius for point-like sources with \suzaku , see also Fig.~\ref{fig-mosaic}). 

However, a larger scale emission component comparable to the intrinsic size of \hj\ may not be detectable if the surface flux in X-rays is sufficiently low. 
For a very conservative estimate of the highest possible X-ray flux from the TeV emission region, we use the ratio between the intrinsic size of \hj\ and the area used for spectral extraction from \csrcTwo\ in this work ($\rho_\mathrm{area} = \left(A_\mathrm{TeV} / A_\mathrm{X-ray}\right)\approx 80$), as well as the ratio of the X-ray surface flux from source-free regions and the region of \csrcTwo\ ($\rho_\mathrm{flux} = \left(F^\mathrm{surf}_\mathrm{source-free} / F^\mathrm{surf}_\mathrm{Suzaku\,J1507.0-6214}\right)\approx 0.33$). 
Assuming that all the diffuse X-ray surface flux from source-free regions is related to \hj\ a scaling factor of $\rho_\mathrm{area} \cdot \rho_\mathrm{flux} \approx 27$ needs to be applied to the measured flux from \csrcTwo . 
The implications of the increased X-ray flux on the model parameters would be an enhanced magnetic field (by a factor of $\sqrt{27} \approx 5$, i.e. $B \approx 2.9\,\mu$G) and a cut-off of the electron spectrum decreased by the same factor, i.e. $E_\mathrm{max} \approx 200$\,TeV \citep[following][]{hinton2009}.

\subsection{Comparison to measured values of the neutral hydrogen column}
\label{sec-nh-measurements}
The values of \nh\ derived from the X-ray spectra in sect.~\ref{sec-analysis} can be compared to measured values of the total absorbing CD of neutral hydrogen in the direction of the sources. 
Soft X-rays are absorbed by interstellar gas, which predominantly consists of atomic (HI) and molecular hydrogen (H$_2$). 

The HI distribution in the galaxy has been measured by \citet{dickey1990} and
\citet{kalberla2005}. 
For the direction of \hj\ \citet{dickey1990} estimated a HI CD of $5.0 \times 10^{21}$~cm$^{-2}$ and \citet{kalberla2005} constrained it to $4.2 \times 10^{21}$~cm$^{-2}$\footnote{http://heasarc.gsfc.nasa.gov/cgi-bin/Tools/w3nh/w3nh.pl}.
In comparison to these values \ssrcOne\ and \csrcTwo\ seem to
be absorbed on a comparable or even higher level. 
This indicates that these sources are located at the edge or even outside the main HI distribution of the Galaxy. 
The excess absorption for the two sources in comparison to the values found in the surveys could in principle be attributed to Galactic molecular hydrogen that is not measured in the HI surveys.

Absorption by neutral hydrogen in soft X-rays is tightly correlated to dust absorption in the optical \citep{predehl1995}. 
Consequently, from the measured level of dust absorption (A$_\mathrm{V}$),
the total CD in cold hydrogen can be estimated. 
For \csrcTwo\ (located inside \hj ) we find A$_\mathrm{V} = 4.6$ and for \ssrcOne\ A$_\mathrm{V} = 3.7$ \citep{schlafly2011}\footnote{http://ned.ipac.caltech.edu/}. 
These measured values of A$_\mathrm{V}$ correspond to a cold hydrogen absorption of $8.2 \times 10^{21}$~cm$^{-2}$ for \csrcTwo\ and $6.6 \times 10^{21}$~cm$^{-2}$ for \ssrcOne\ \citep{predehl1995}.
These comparisons evidence that both sources experience soft X-ray absorptions
consistent with the total Galactic CD of neutral hydrogen. 
However, it has to be noted that the measurement of dust absorption may not be precise at Galactic latitudes $< 5 \degr$ \citep{schlafly2011}.

An alternative method to determine the total X-ray absorbing hydrogen CD is described by \citet{willingale2013} who used 493 gamma-ray burst afterglows to calibrate estimates of \nh\ based on measurements of atomic hydrogen and dust extinction. 
Using their web tool\footnote{http://www.swift.ac.uk/analysis/nhtot/index.php} to estimate the total ($N_\mathrm{HI} + N_\mathrm{H2}$) for the region of \hj\ we get a value of \nh\~=~5.7$\cdot 10^{21}$\,cm$^{-2}$ {and for \ssrcOne we obtain \nh\~=~5.9$\cdot 10^{21}$\,cm$^{-2}$. Both values are slightly lower than estimated from the dust absorption A$_\mathrm{V}$ directly (see previous paragraph).

\subsection{Comparisons to models of the Galactic neutral hydrogen distribution}
\label{sec-nh-model}
The highly absorbed X-ray spectrum of \csrcTwo\ may suggest that any associated object is located at a considerably large distance to Earth. 
In order to estimate the source distance, we use a large-scale 3D model for both atomic and molecular Galactic hydrogen gas and compare it to the CD derived from the \suzaku\ measurements. 
This model is based on the descriptions of the large-scale distribution of HI and $\mathrm{H}_\mathrm{2}$ gas in the Milky Way by various authors, summarized by \citet{ferriere2001}. 
There, a galacto-centric distance of the sun of $R_\odot$ = 8.5\,kpc is assumed, and gas distributions that apply different values of $R_\odot$ are rescaled to the updated value by the author. 
We also adopt this value of $R_\odot$ in the following. 

The large-scale distribution of each of the mentioned gas components is modeled as follows: First, the radial (i.e. with $R$, the distance to the Galactic center along the Galactic plane) distribution of the azimuth averaged gas CD \textit{perpendicular} to the Galactic plane (vertical CD) is obtained. Second, the radial profile of the gas layer thickness is determined. Lastly, a vertical density distribution function is assumed, and its normalization and width is determined by the two other distributions. As a result, for each set of coordinates $\lbrace R,z \rbrace$, where $z$ is the vertical distance to the Galactic plane, a space-averaged number density of hydrogen nuclei can be calculated.

Furthermore, we allow for a large-scale Galactic spiral arm structure by modulating the gas densities towards the arms in a simple manner, as described in sect.~\ref{sec-spiral-arms}.

\subsubsection{Atomic hydrogen (HI)}
\label{sec-Hone}
\citet{dickey1990} provide the azimuth averaged radial distribution of the HI vertical CD, which  can be roughly described by a three-component structure: a depletion at the Galactic center followed by an increase to a value of $6.2\times 10^{20}\mathrm{cm}^\mathrm{-2}$ at a distance of $R = 3.5$\,kpc (region (i)). This transitions into a constant regime (region (ii)) which passes through the solar circle and out to almost  $R = 14$\,kpc. At farther distances (region (iii)), the values of the vertical CD again decrease in an exponential fashion. We use the following parameterization to describe the radial distribution of the vertical HI CD:
\begin{equation*}
\label{CDHIone}
\mathrm{CD}_\mathrm{HI} =(6.2\cdot10^{20}\mathrm{cm}^\mathrm{-2} )\begin{cases} \exp{\left\{-\left[(R-3.5)/0.7\right]^2\right\}} &\mathrm{(i) } \\
           1  &\mathrm{(ii) }\\
           \exp\left\{-\left(R-13.65\right)/3.57\right\} &\mathrm{(iii), }

       \end{cases}
\end{equation*}
with $R$ in units of kpc.

The HI gas layer thickness is modeled according to \citet{dickey1990}: The FWHM increases from the Galactic center outwards to $R = 3.5$\,kpc (region (a)) from 165\,pc to 230\,pc. Between $R = 3.5$\,kpc and the solar circle (region (b)) the layer thickness is assumed to remain constant at a value of 230\,pc. We model the flaring of the gas layer outside the solar circle (region (c)) by a linear increase to a value of about 3 kpc at $R = 25$\,kpc.

\begin{equation*}
\label{CDHItwo}
\mathrm{FWHM}_\mathrm{HI} =(1\mathrm{ kpc})\begin{cases} 0.065\exp{\left\{-\left[(R-3.5)/7.1\right]\right\}}+0.165 &\mathrm{(a) } \\
          0.23 &\mathrm{(b) }\\
          0.154\cdot(R-8.5)+0.23 &\mathrm{(c) }

       \end{cases}
\end{equation*}

Finally, we use the vertical density profile reported by \citet{ferriere2001}, resembling two Gaussian functions plus an exponential tail: 

\begin{eqnarray*}
\label{HIzProfile}
\begin{aligned}
\langle n_{\mathrm{HI}}\rangle(R,z) &=&k(R)\left[0.7\cdot \exp\left\lbrace-\left(\frac{z}{0.55\cdot \mathrm{FWHM}_\mathrm{HI}(R)}\right)^2\right\rbrace\ \right. \\
  & &+0.19\cdot \exp\left\lbrace-\left(\frac{z}{1.38\cdot \mathrm{FWHM}_\mathrm{HI}(R)}\right)^2\right\rbrace\ \\
  & &\left. +0.11\cdot \exp\left\lbrace-\left(\frac{|z|}{1.75\cdot \mathrm{FWHM}_\mathrm{HI}(R)}\right)\right\rbrace\right]\mathrm{cm}^\mathrm{-3}.
\end{aligned}
\end{eqnarray*}

Here we are rescaling the width of the function according to our model of the HI gas layer thickness while the normalization factor $k(R)$ is fixed by $\mathrm{CD}_\mathrm{HI}(R)$.

\subsubsection{Molecular hydrogen (H$_2$)}
\label{sec-htwo}
We model the distribution of the molecular hydrogen component in a similar way, using the azimuth-averaged results of \citet{clemens1988} for H$_2$ in the first Galactic quadrant.
The radial distribution of the vertical molecular hydrogen CD is modeled empirically to follow the distribution provided by \citet{clemens1988}\footnote{Rescaled to the updated value of $R_\odot = 8.5$\,kpc, see \citet{ferriere2001}.} as
\begin{equation*}
  CD_{\mathrm{H}_\mathrm{2}} = (1.54\cdot 10^{19}\mathrm{ cm}^\mathrm{-2}) \cdot R^{8}\exp\lbrace-1.69\cdot R\rbrace.
\end{equation*}

For the FWHM of the $\mathrm{H}_\mathrm{2}$ layer thickness as function of the Galacto-centric radius, we use the powerlaw parameterization by \citet{clemens1988}: 

\begin{equation*}
  \mathrm{FWHM}_{\mathrm{H}_\mathrm{2}} = \left(562 \cdot R\right)^{0.58} \textrm{pc}.
\end{equation*}

The density distribution perpendicular to the Galactic plane follows the shape described in \citet{ferriere2001}, again rescaled in width and normalization $l(R)$ as given by $CD_{\mathrm{H}_\mathrm{2}}(R)$ and
$\mathrm{FWHM}_{\mathrm{H}_\mathrm{2}}(R)$,

\begin{equation*}
  \langle n_{\mathrm{H}_\mathrm{2}}\rangle(R,z)  = l(R) \exp\left[-4 \ln 2 \left(\frac{z}{\mathrm{FWHM}_{\mathrm{H}_\mathrm{2}}(R)}\right)^2\right] \text{cm}^\text{-3}.
\end{equation*}

\subsubsection{Spiral Arms}
\label{sec-spiral-arms}
As an option, we also allow for the modulation of the gas density towards Galactic spiral arms.
In the inter-arm regions \citet{clemens1988} found values of the space-averaged density of molecular hydrogen lowered by a factor of $\sim3.6$ compared to the density inside the arms. A similar contrast of $\sim4$ is found for the HI surface density, see \citet{kulkarni1982}.

The geometrical arm model is taken from \citet{valee2005}, but we rescaled the spirals to match our value of $R_\odot$ = 8.5\,kpc, which is somewhat larger than Val\'ee's value of 7.9\,kpc.
The width of the spiral arms is adopted from \citet{russeil2003}, who found a FWHM value for all four arms of $\langle w \rangle$  = 1.32\,kpc, corresponding to a Gaussian width of $\sigma_w$ = 0.56\,kpc.
The space-averaged gas density follows a Gaussian profile perpendicular to the spiral arm tangential direction so that the density between the arms is a factor of 4 lower than in their center,
\begin{equation*}
\langle n_{mod} \rangle = (\langle n_{\mathrm{HI}}\rangle + \langle n_{\mathrm{H}_\mathrm{2}}\rangle) \cdot s\cdot \left[ 3 \exp{(-d^2/2 \sigma_w^2)} +1 \right].
\end{equation*}
Here, $d$ is the distance to the nearest spiral arm and $s$ = 0.32 is an empirically determined normalization factor to obtain the same
number of hydrogen nuclei in both the modulated and unmodulated models. 

The resulting gas densities, averaged perpendicular to the Galactic plane, are shown in Fig. \ref{fig-galaxy}.

\begin{figure}
  \begin{center}
  \resizebox{1.0\hsize}{!}{\includegraphics[clip=]{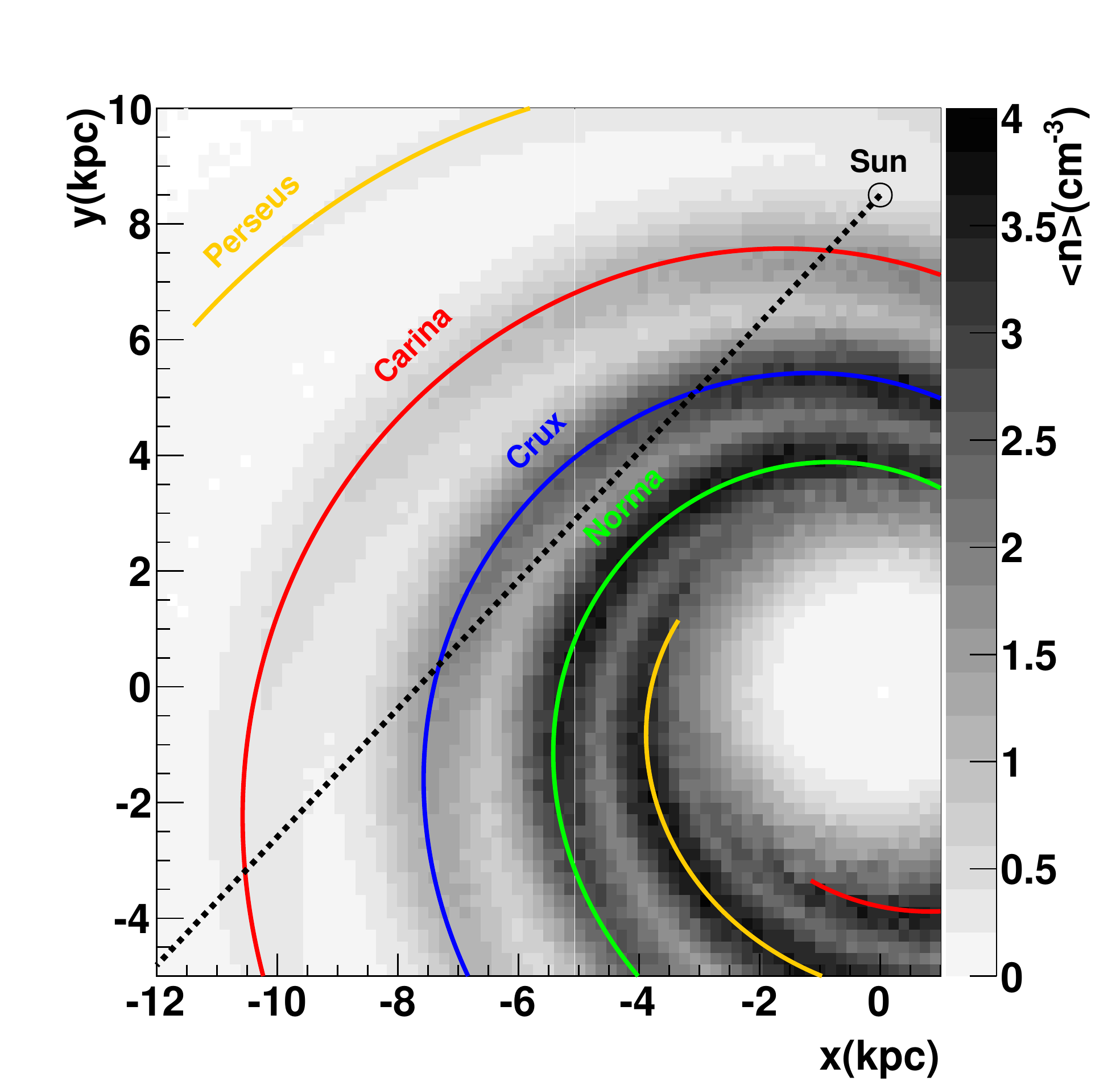}}  
  \caption{$\mathrm{HI}+\mathrm{H}_\mathrm{2}$ atomic nucleus density, averaged along the perpendicular direction to the Galactic plane. Additionally, the line of sight towards \hj\ (dashed) and the Val\'ee spirals (colored) are shown.}
  \label{fig-galaxy}
  \end{center}
\end{figure}

\begin{figure*}
  \begin{center}
  \resizebox{1.0\hsize}{!}{\includegraphics[clip=]{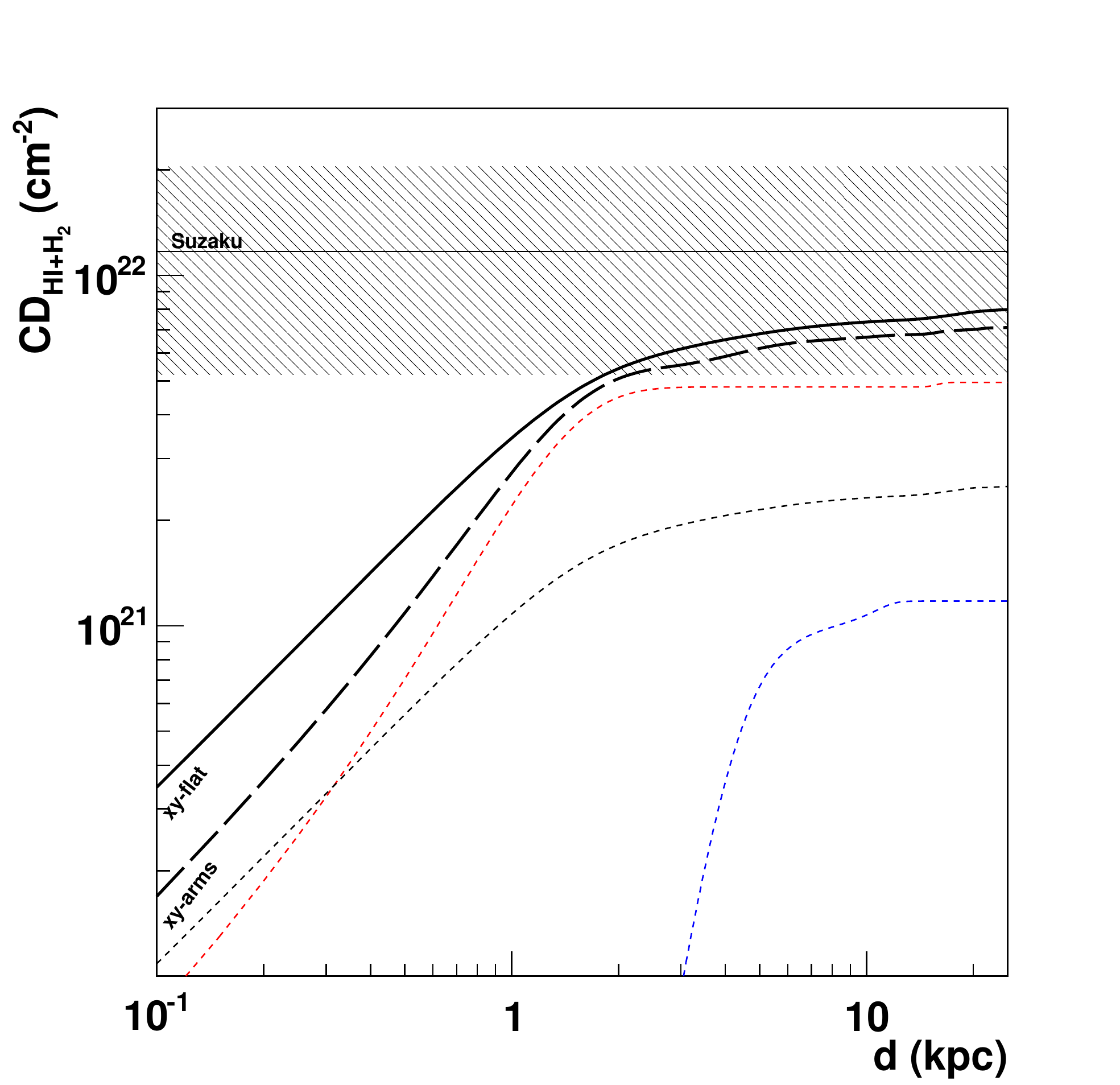}
  	\includegraphics[clip=]{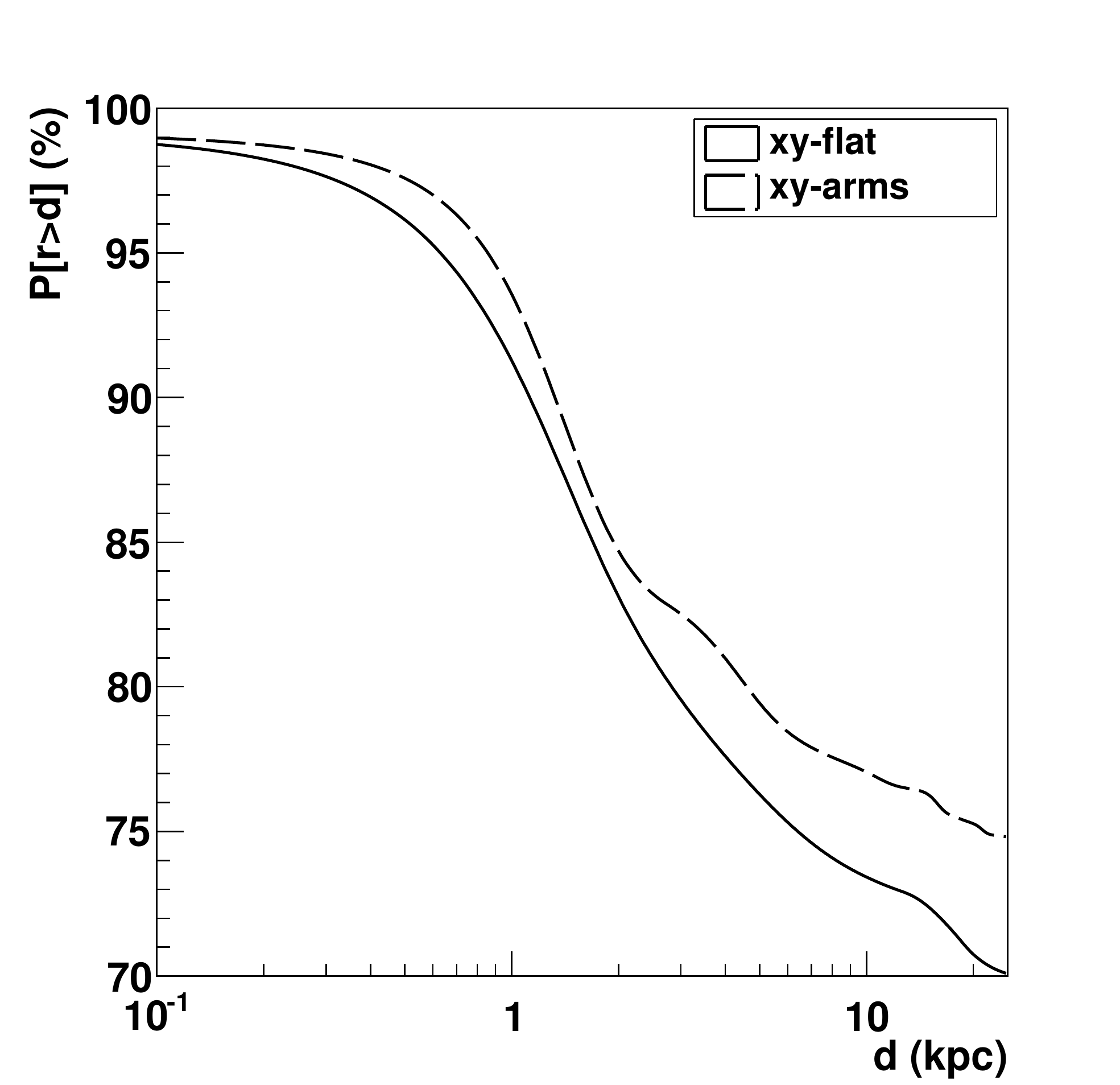}}
  \caption{\textbf{Left:} Calculated CD of HI and $\mathrm{H}_\mathrm{2}$ gas in the direction of \hj\ as a function of the distance to Earth. The solid line shows the model expectation in the absence of any Galactic spiral arm gas modulation while the thick dashed line takes such a modulation into account. The thin dashed lines represent the contributions of the inter-arm (black) and the intra-arm gas (colored, see Fig. \ref{fig-galaxy}). \textbf{Right:} Probability profile as a function of the distance to Earth for \csrcTwo\ being located at a distance $r$ greater than $d$.}
  \label{fig-model1}
  \end{center}
\end{figure*}

\subsection{The distance to \hj}
\label{sec-distance}
The model described in sect.~\ref{sec-nh-model} allows to derive the theoretically expected neutral hydrogen CD in the direction of \csrcTwo\ as a function of the distance to Earth, $d$. 
The result is shown in the left panel of Fig.~\ref{fig-model1}. For distances larger than $\sim$2\,kpc, both models agree within errors with the \suzaku\ measurement. The spiral arm modulation results in overall lower values for $d$ than those obtained for a 'flat' galaxy in the x-y plane. This is especially pronounced at distances smaller than 1\,kpc, as the main contribution to the CD stems from the gas concentrated in the Carina arm (red line in Fig. \ref{fig-model1}, see also Fig. \ref{fig-galaxy}).

We use the modeled CD profile along $d$ to calculate the probability $P[r>d]$ of \csrcTwo , and presumably \hj , being located at a distance $r > d$. 
To that end, we treat the asymmetric error interval
of the X-ray \nh\ measurement as a distorted Gaussian distribution, following Method 2 described in \citet{barlow2004}. 
The resulting profile is shown in Fig.~\ref{fig-model1}. As can be seen, the X-ray measurements and Galactic gas model places \csrcTwo\ with probabilities of 70\% (flat galaxy) or 75\% (arm-modulated) at distances larger than 25\,kpc, which corresponds to the edge of the Galaxy. 
Interestingly, \hj\ is located very close to the super-galactic plane (SG-lon: 183.6$^\circ$, SG-lat: 0.2$^\circ$), a location where nearby Extragalactic
objects cluster \citep{lahav2000}, and also where all non-blazar Extragalactic TeV sources are concentrated (see TeVCat\footnote{http://tevcat.uchicago.edu/}). 
Although interesting, these values, however, do not yet allow to make any definitive statements on the exact location of \hj , due to the relatively large uncertainties of the \nh\ measurement from the X-ray spectrum. 

Our constraints on the distance to \hj\ can be compared to the distance estimates by \citet{matsumoto2014} for \ssrcOne\ of $d = 1.7 \pm 0.2$\,kpc and for \csrcTwo\ of $d = 2.6 \pm 1.7$\,kpc. 
We note here that also their measurements of the total \nh\ are compatible with the total Galactic CD within 1\,$\sigma$ uncertainties, and thus the calculation of an upper bound of the distances to these two sources should not be possible, contrary to what is indicated by the symmetric distance uncertainties quoted by \citet{matsumoto2014}. 
In such cases, the only constraints on the distance can be given in terms of probabilities for $d$ to be larger than a certain value $r$, as, e.g., derived from our model (see Fig.~\ref{fig-model1}). 
Furthermore, the estimates by \citet{matsumoto2014} are based on the assumption of an average Galactic ISM density of 1\,cm$^{-3}$, which is probably not appropriate for the specific case of \hj , due to its large offset from the Galactic plane and hence significantly lower ambient densities, particularly for large distances. 
This assumption of a too dense medium systematically biases the resulting distance estimates towards lower values. 

We note here that the total values of \nh\ derived from the Galaxy model are comparable or slightly larger than those estimated from measurements (see sect.~\ref{sec-nh-measurements}.
Therefore, the derived probabilities of \hj\ being located at a distance $r>d$ in this section can be viewed as conservative, since lower values of \nh\ in the model would result in even larger distances. 

For comparison, we apply the same procedure to \ssrcOne . 
The result can be seen in Fig. \ref{fig-modelsource2}. 
Both the unmodulated and spiral arm model predictions for the absolute neutral hydrogen CD in the direction of the source agree within errors with the \suzaku\ measurements. 
As the predicted total hydrogen CD in case of the modulated gas model almost perfectly coincides with the X-ray value, the estimated probability for \ssrcOne\ to be extragalactic or at the edge of the milky way is $\sim$50\% in this case. 
With the unmodulated gas model, the total hydrogen CD is predicted at a higher value but still matches the upper 1$\sigma$-limit of the X-ray measurement, favoring a Galactic nature of the source with a probability of $\sim$84\%. Again, our model predictions on the total hydrogen CD are somewhat larger than the independent measurements presented in sect.~\ref{sec-nh-measurements}, which would result in lower probabilities of \ssrcOne\ being a Galactic source.

\begin{figure*}
  \begin{center}
  \resizebox{1.0\hsize}{!}{\includegraphics[clip=]{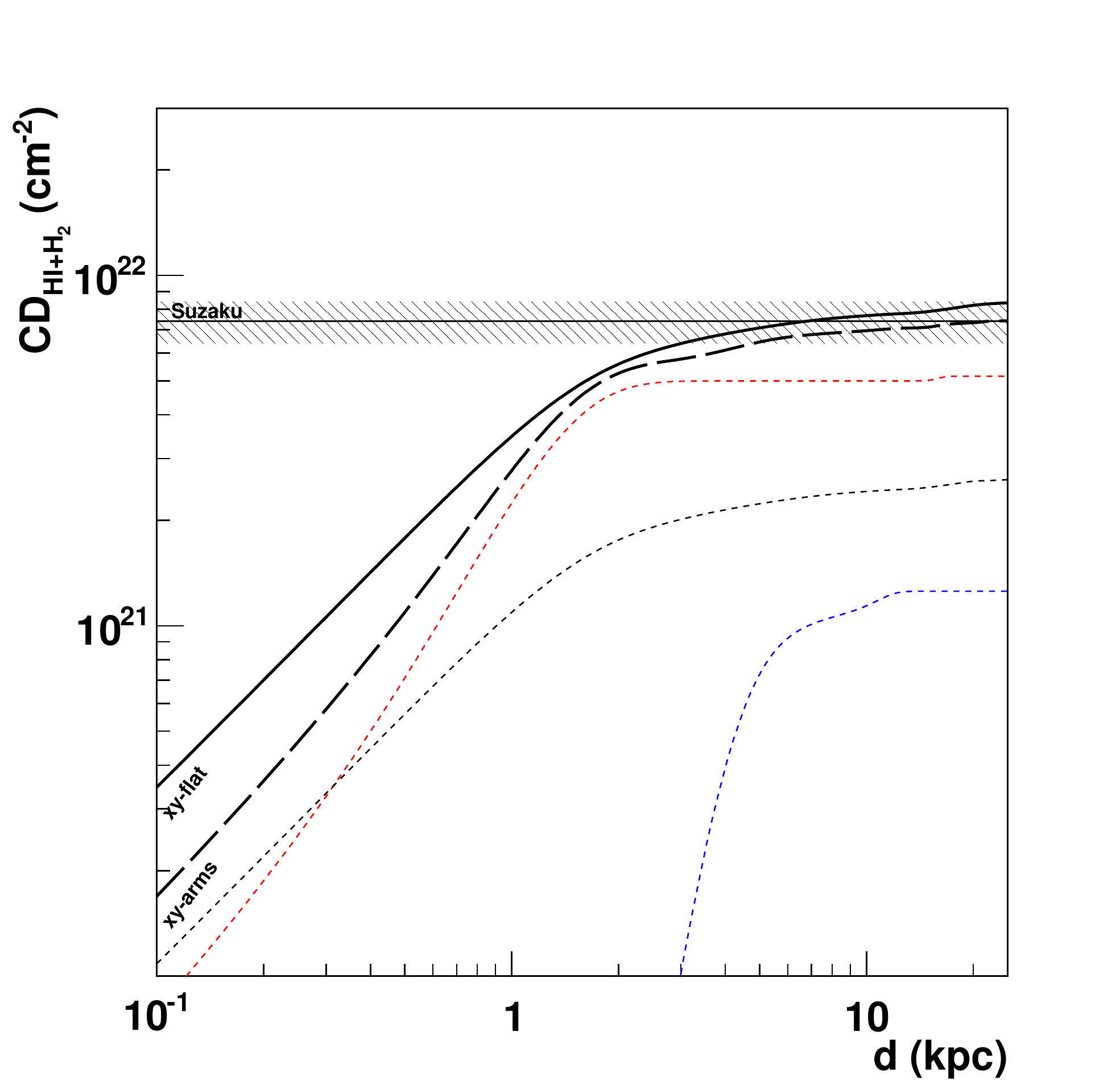}
  	\includegraphics[clip=]{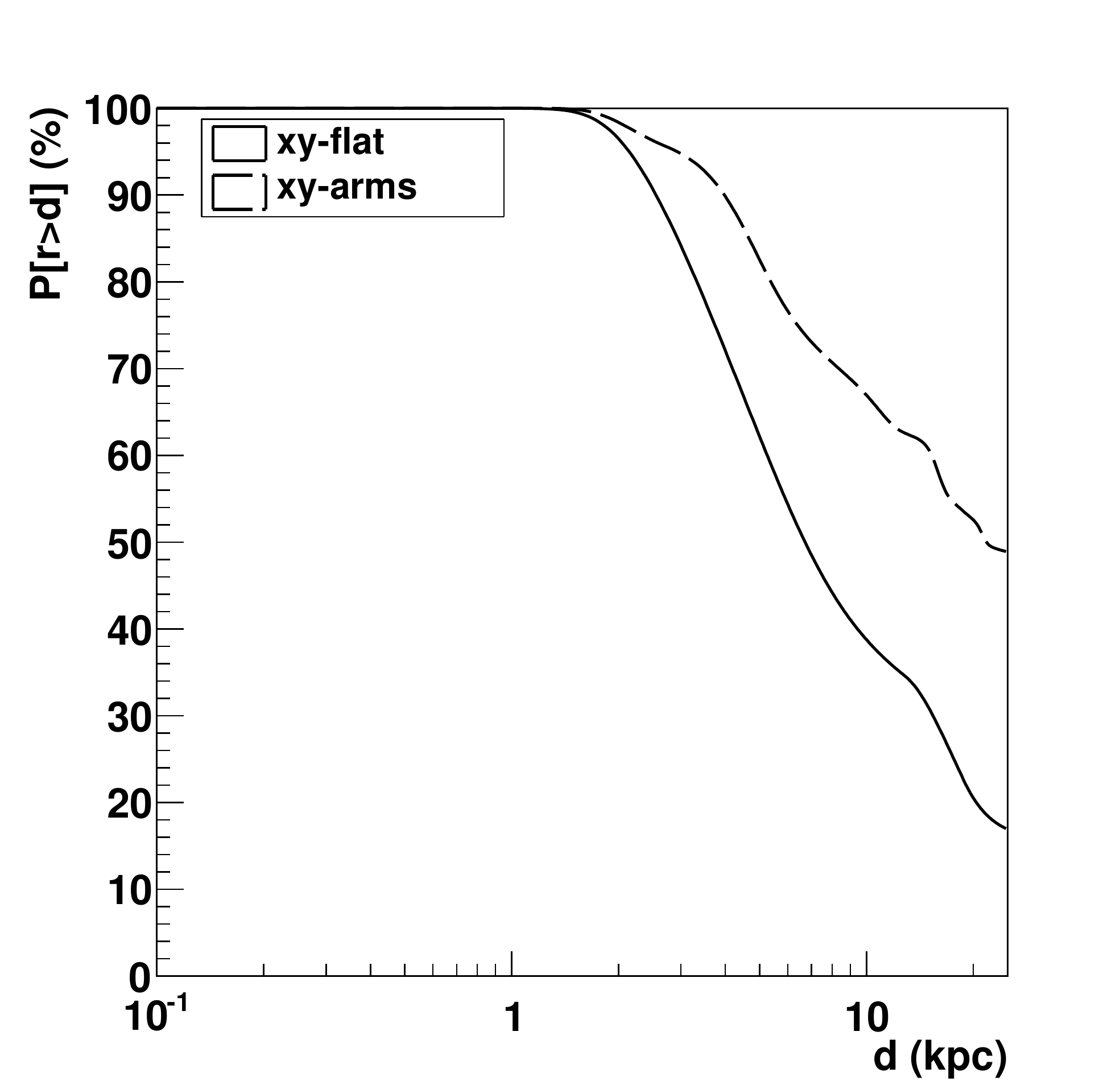}}
  \caption{Same as in Fig.\ref{fig-model1}, but calculated for \ssrcOne .}
  \label{fig-modelsource2}
  \end{center}
\end{figure*}

\subsection{Implications for a PWN scenario}
\label{sec-pwn-scenario}
As discussed in the previous section (\ref{sec-distance}) the comparatively large value of \nh\ measured in X-rays might indicate a considerable line-of-sight distance of \csrcTwo , and potentially \hj , from Earth. 
Given the rather high Galactic latitude ($b=-3.49^\circ$), this would also imply a large vertical distance to the Galactic plane, $z$. 
Also, the Galactic longitude ($b=317.97^\circ$) of \hj\ defines a minimal value of $R$, $R_{min}$ = 5.7\,kpc. 
These distances, together with a model for the distribution of PWNe in the Galaxy, are used in the following to estimate the probability of a PWN scenario for \hj .

\subsubsection{Un-recycled pulsars}

First we investigate the probability of \hj\ being the PWN associated with an un-recycled pulsar.
This is the population of pulsars considered in the study of \citet{mattana2009}.
We assume that PWNe follow the distribution of pulsars in the Milky Way. Corresponding radial and vertical distribution functions have been proposed by \citet{yusifov2004} and \citet{faucher2006}, respectively, and are adopted in our model. 
The mentioned radial distribution describes a powerlaw increase followed by an exponential decay, $\rho(R) \sim R^{1.64} \exp \left\lbrace -4.01\cdot (R-R_\odot)/R_\odot\right\rbrace$, and reaches its maximum at $R = 3.2$\,kpc. 
For the distribution perpendicular to the Galactic plane we assume an exponential profile with scale height $z_0$. 
The value of $z_0$ is estimated by a fit to the $z$-distribution of
pulsars detected by the Fermi-LAT instrument \citep[][the Second \emph{Fermi} Large Area Telescope Catalog of Gamma-Ray Pulsars]{abdo2013}, 
which provides a homogeneous sky coverage.
To fit the pulsar distribution we used all Galactic Fermi-detected un-recycled pulsars without any cut on age or spin-down power.
This results in a sample of 51 pulsars. 
An exponential fit to the $z$-distribution ($\chi / ndf = 9.6/7$) yields a scale height of $z_0=(74.0\pm16.1)$\,pc. 

With this distribution function one can determine the probability $\tau$ to find a pulsar at any given coordinate $\lbrace R>R_{min}, z(\tau) \rbrace$.
For instance, assuming a value of $z_0=74$\,pc, at a perpendicular distance to the Galactic plane of $z(0.001\%)$ = 815\,pc, the chance to find a pulsar is $\tau$ = 0.001\%. 
We calculated the CD towards \hj\ that corresponds to $z(0.001\%)$ and compare it to the X-ray measurement of \nh , again treating the probability distribution of the latter as a distorted Gaussian (see Sec.~\ref{sec-distance}). 
This comparison places the sources with a probability of 73\% (flat galaxy) and 76\% (spiral arms) at values $z > z(0.001\%)$. 
Thus, at a confidence of 99.999\%, model and measurement exclude a PWN scenario for \hj\ with probabilities between 73\% and 76\%.
Probabilities for different values of $\tau$ are given in Tab.~\ref{PWNprob2} and also shown in Fig.~\ref{fig-model2} for $z_0=74$\,pc.
Again, given the relatively large uncertainties of the X-ray measurement of \nh , these results do not allow to disfavor a PWN scenario with high confidence.

\begin{figure}
  \begin{center}
\resizebox{0.97\hsize}{!}{\includegraphics[clip=]{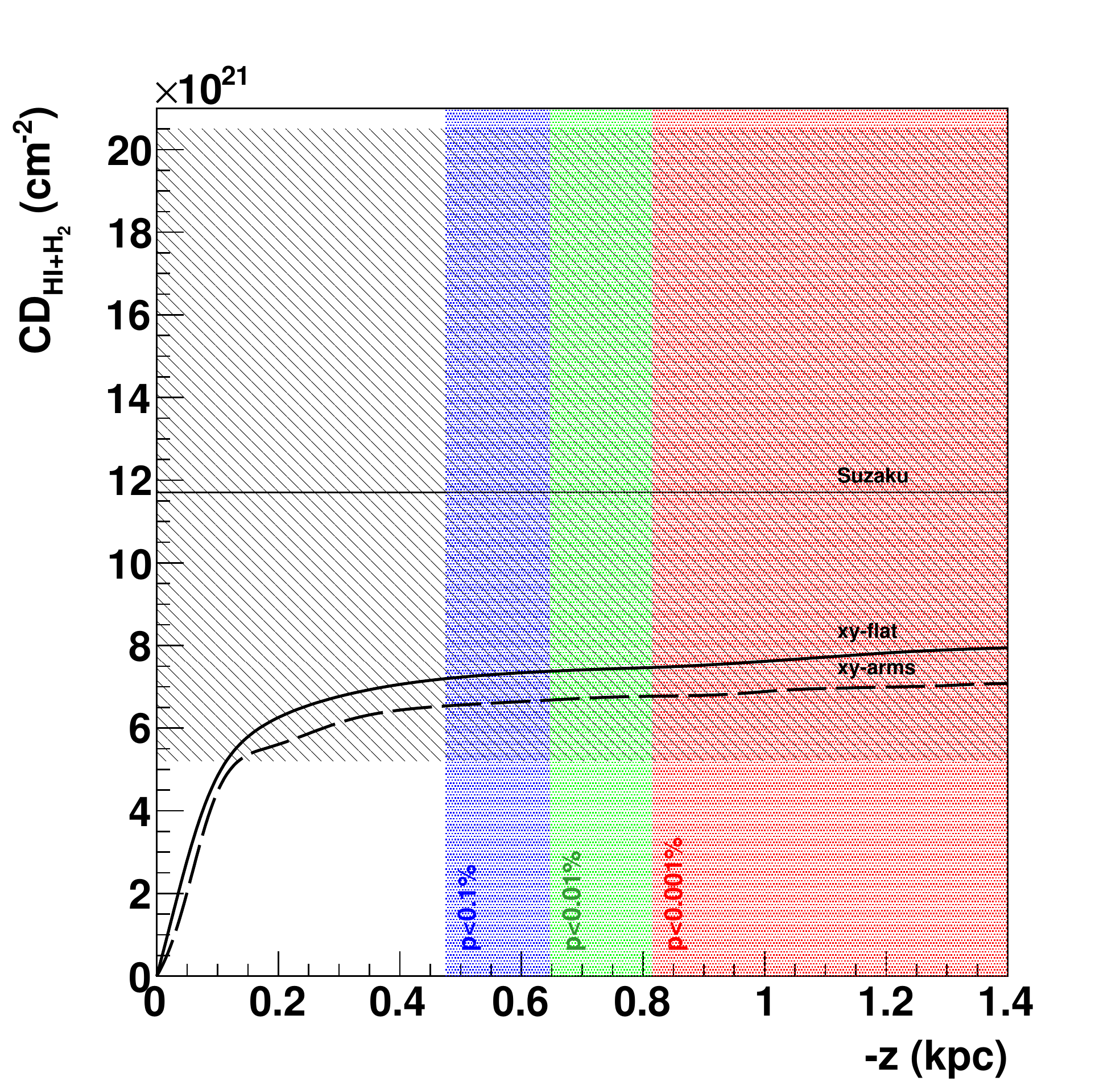}}  
  \caption{Same as in the left panel of Fig. \ref{fig-model1}, but here as a function of $z$. Additionally, the $z$-values corresponding to $\tau=0.1\%, 0.01\%, 0.001\%$ and assuming a vertical scale height of the pulsar distribution of $z_0=74$\,pc are shown.}
  \label{fig-model2}
  \end{center}
\end{figure}

\begin{table}
\begin{center}
\caption[]{Non-PWN scenario probability as a function of $z$. Probabilities have been calculated for the exponential scale height of the pulsar distribution 
$z_0$ of 74\,pc.}
\renewcommand{\tabcolsep}{4pt}
\begin{tabular}{l|lll}
\hline
\hline
$\tau$ & $z(\tau)/\mathrm{pc}$ & $P_{flat}$ & $P_{arms}$\\
\hline
0.1\% &  475 & 74\% & 78\% \\
0.01\% &  647 & 73\% & 77\% \\
0.001\% &  815 & 73\% & 76\% \\
\hline
\end{tabular}
\label{PWNprob2}
\end{center}
\end{table}

\subsubsection{Millisecond pulsars}

In this section we discuss the implication of the rather large value of \nh\ for the case where \hj\ is the nebula associated with a millisecond pulsar (MSP).
MSPs are commonly believed to be very old pulsars with characteristic ages of 10$^9$~years that were recycled
by accretion from a companion star \citep[e.g.][]{alpar1982,gregoire2013}. 
Following the discussion of the potentially old age of \hj\ in Sect.~\ref{sec-sed-model}, a MSP may be a potential counterpart.
Owing to their age their distribution features a large Galactic scale height of about 1\,kpc and a Galactic scale length of about 4\,kpc 
\citep[as measured from the \emph{Fermi-LAT} detected MSPs][]{gregoire2013}.
This scale height is larger than the scale heights of cold hydrogen and therefore no constraints on the probability of \hj\ being a PWN associated to a MSP can be made. 
However, so far no detection of a PWN related to a MSP in VHE gamma-rays has been reported in the literature. 
However, this can in principle be a selection bias since MSPs do not strongly cluster around the Galactic plane, the region with the deepest large-scale exposure by VHE gamma-tray telescopes such as H.E.S.S.
Future observations of powerful MSPs are required to test the presence of VHE-gamma-ray emitting PWNe around these objects. 

\section{Summary and Outlook}
\label{sec-summary}
In this paper we present spectral results for the potential X-ray counterpart to the enigmatic TeV gamma-ray source \hj .
On the basis of this new measurement we provide interesting constraints on the parameters of the underlying population of relativistic particles, and also on the line-of-sight distance to the object. 
Even though this study represents a significant step forward in the identification of the origin of non-thermal emission from this puzzling object, the relatively low count statistics of the X-ray detection, and the resulting large uncertainties of the derived model parameters do not allow very strong conclusions. 
However, we see indications that the potential X-ray counterpart to \hj\ might be located at a considerable distance from Earth, which would strongly challenge established models for Galactic gamma-ray source populations, in particular PWNe. 

Future deeper X-ray observation with high-throughput instruments providing 
higher statistical quality could greatly improve the strength and conclusiveness of this result.

\section*{Acknowledgments}
This research has made use of data obtained from the Suzaku satellite, a collaborative mission between the space agencies of Japan (JAXA) and the USA (NASA).  This research has made use of data and/or software provided by the High Energy Astrophysics Science Archive Research Center (HEASARC), which is a service of the Astrophysics Science Division at NASA/GSFC and the High Energy Astrophysics Division of the Smithsonian Astrophysical Observatory.

\label{lastpage}

\end{document}